\documentclass{aa}
\usepackage{graphicx}
\usepackage{txfonts}
\usepackage{natbib}

\begin{document}

\newcommand{\um}{\,$\mu$m}

\title{Carriers of the mid-IR emission bands in PNe reanalysed.}
\subtitle{Evidence of a link between circumstellar and interstellar aromatic dust
\thanks{This work is based on observations made with the Spitzer Space
Telescope, which is operated by the Jet Propulsion Laboratory, California Institute of Technology 
under a contract with NASA.
Based on observations with ISO, an ESA project with instruments funded by ESA Member States
(especially the PI countries: France, Germany, the Netherlands and the United Kingdom) and
with the participation of ISAS and NASA.}}

\author{C. Joblin\inst{1}
\and 
R. Szczerba\inst{2}
\and
O. Bern\'e \inst{1}
\and
C. Szyszka\inst{2,3}
}

\offprints{\\ C.~Joblin \email{christine.joblin@cesr.fr}}
\institute{
Centre d'Etude Spatiale des Rayonnements, Universit\'e de Toulouse et CNRS,
Observatoire Midi-Pyr\'en\'ees, 9 Av. du Colonel Roche, 
31028 Toulouse cedex 04, France
\and
N. Copernicus Astronomical Center, Rabianska 8, 87-100 Toru\'n, Poland 
\and
Nicolaus Copernicus University, Gagarina 11, 87-100 Toru\'n, Poland 
}

\date{Received ?; accepted ?}

\abstract
{It has been shown that the diversity of the aromatic emission features can be rationalized into different
classes of objects, in which differences between circumstellar and interstellar matter
are emphasised. 
}
{ We probe the links between the mid-IR emitters observed in planetary nebulae (PNe)
and their counterparts in the interstellar medium in order to probe a scenario in which the latter
have been formed in the circumstellar environment of evolved stars. 
}
{The mid-IR (6-14\um) emission spectra of PNe and compact \ion{H}{ii} regions were analysed
on the basis of previous work on photodissociation regions (PDRs).
Galactic, Large Magellanic Cloud (LMC), and Small Magellanic Cloud (SMC) objects
were considered in our sample.
}
{We show that the mid-IR emission of PNe can be decomposed as the sum of six components.
Some components made of  polycyclic aromatic hydrocarbon (PAH) and very small grain (VSG) populations
are similar to those observed in PDRs. Others are fitted in an evolutionary scenario involving
the destruction of the aliphatic component observed in the post-AGB stage, as well as strong
processing of PAHs in the extreme conditions of PNe that leads to a population of very large ionized PAHs.
This species called PAH$^x$ are proposed as the carriers of a characteristic  band at 7.90\um.
This band can be used as part of diagnostics that identify PNe
in nearby galaxies and is also observed in galactic compact \ion{H}{ii} regions. 
}
{These results support the formation of the aromatic very small dust particles in the envelopes of
evolved stars, in the Milky Way, as well as in the LMC and SMC, and their subsequent survival
in the interstellar medium.
}


\authorrunning{Joblin et al.}
\titlerunning{Linking aromatic dust in PNe and ISM}

\maketitle

\section{Introduction} \label{int}

\begin{figure*}[ht!]
\includegraphics[width=9 cm] {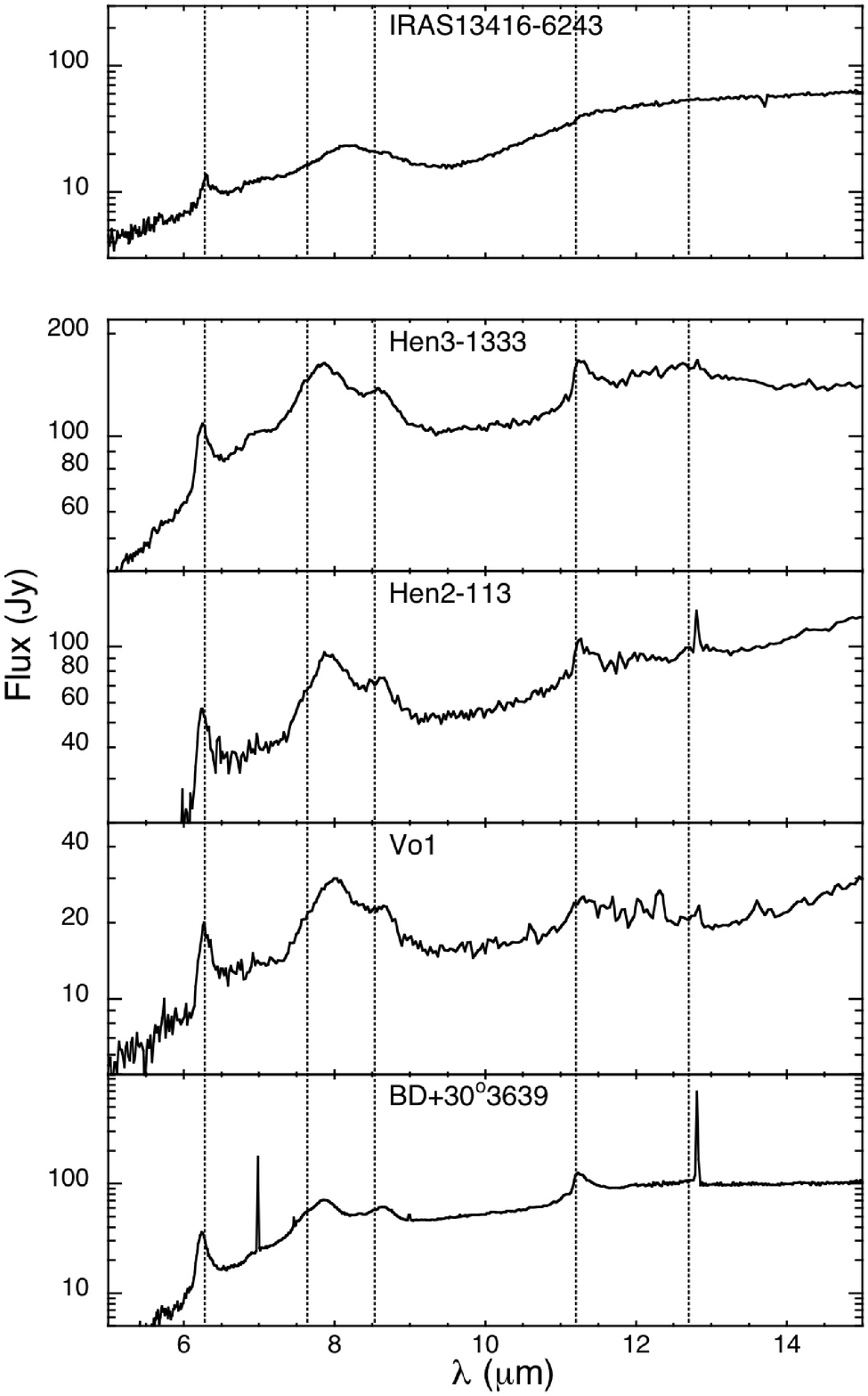}
\includegraphics[width=9 cm]{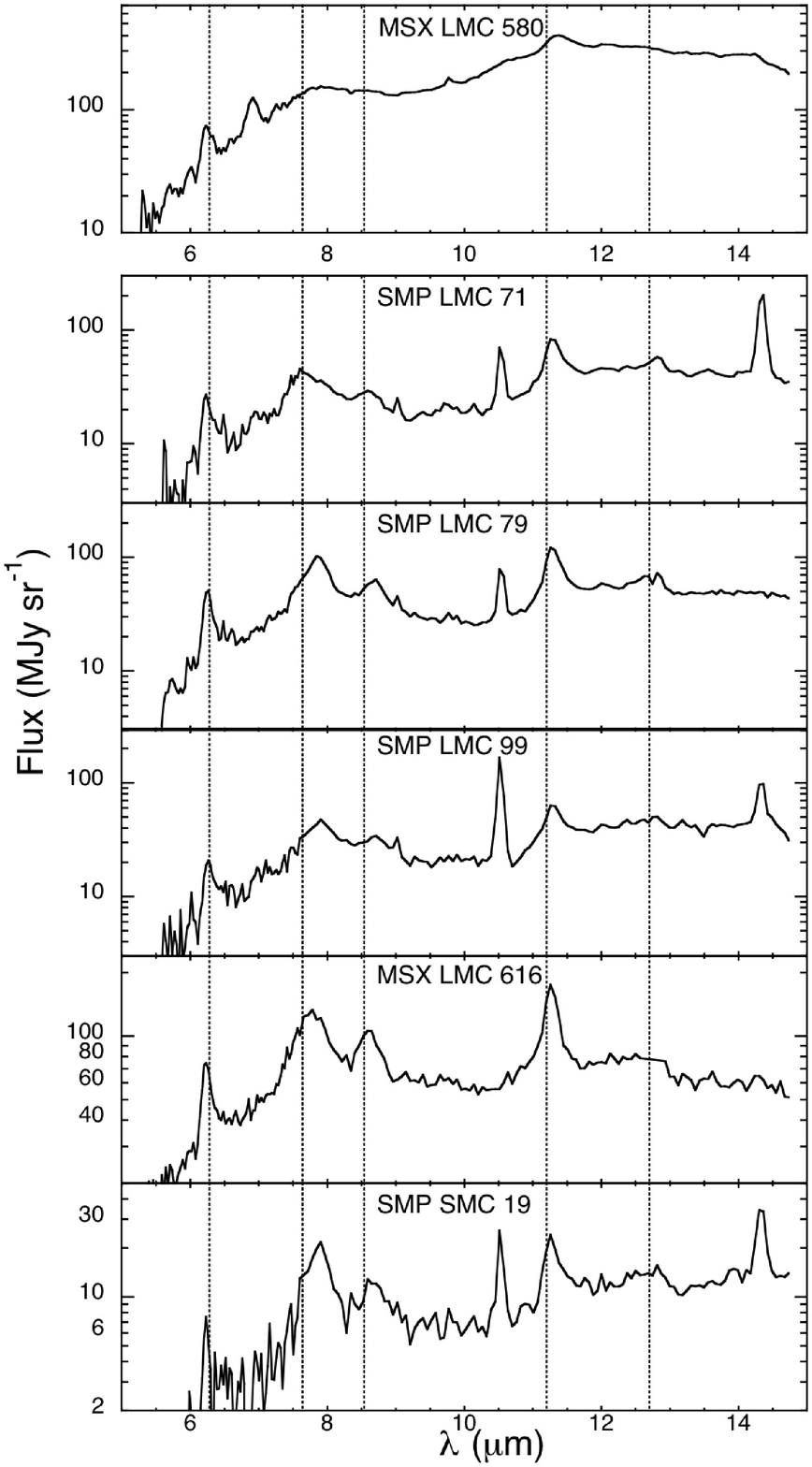}
\caption{Mid-IR (5-15\um) spectra of a post-AGB object (top panel)  and of the studied PNe, in the Milky Way
(left panel; ISO SWS01 observations) and in the LMC/SMC (right panel; Spitzer-IRS observations).
Vertical lines have been drawn at 6.28, 7.64, 8.55, 11.20, 12.70\um, at the peaks of the template
PAH$^{+}$ spectrum (cf. Table\,\ref{tab:temp}).}
\label{fig:rawspec}
\end{figure*}

Evolved stars are known to play an important role in the dust formation in galaxies, but the link
between circumstellar and interstellar dust is still an open question.
Dust populations include the carriers of the so-called
aromatic infrared bands (AIBs), which are located between 3.3 and 14\um\ and 
attributed to very small aromatic dust particles
amongst which are large polycyclic aromatic hydrocarbons (PAHs).
Emission in the AIBs arises following stochastic heating of their carriers by UV photons,
and it is therefore often used to probe UV-excited environments in our Galaxy as well as
in external galaxies \citep{uch00, pee04, bra06, spo07, smi07, leb07}.
A detailed analysis of the band positions and profiles can also be used as
a diagnostics of the type of emitting region  \citep{pee02, van04}.
Ultimately, one would like to use these bands to trace the chemical evolution of their carriers
induced by the local physical conditions.

Several authors have tried to provide a chemical scheme
based on soot formation models to form PAHs in evolved carbon stars \citep{fre89, che92}.  This
scheme, which involves the pyrolysis of hydrocarbons and more especially of C$_{2}$H$_{2}$
has gained support with the detection of  C$_{4}$H$_{2}$,  C$_{6}$H$_{2}$
and  C$_{6}$H$_{6}$ in CRL618 \citep{cer01}.
The study of the AIB carriers in these cool-star environments is difficult, however, due to the few UV
photons available to trigger the AIB emission.
There is therefore a natural bias towards their study in planetary nebulae (PNe) where
the central star is very hot.
However, the AIB spectrum in PNe is known to
present significant differences compared to the bands observed in PDRs.
In particular, the dominant band of ``7.7\um'' feature
arises at 7.8\um\ in PNe compared to  7.6\um\ in the ISM \citep{bre89, coh89, pee02}.
We might therefore suspect a significant chemical evolution
between PAHs in these circumstellar environments and those dominating the emission
at the scale of a galaxy and located in PDRs \citep{hon01}.
To progress on this question, we propose to analyse the spectra of several PNe in the
Milky Way (MW) and in the Large and Small Magellanic Clouds (LMC and SMC)
as examples of galaxies with metallicities that differ by a factor of 2 to 4.

The analysis of the spectra is inspired from previous work by 
\citet{rap05} and \citet{ber07} on cool PDRs.
The authors find that the 7.6\um\ component is carried by molecular
PAHs with two charge states, neutrals and cations, whereas some very small grains (VSGs)
carry the mid-IR continuum and broader bands, in particular one at $\sim$7.8\um.
In Sect. 2, we describe the sample of objects and data reduction. The fitting procedure is detailed
in Sect. 3 and the results are presented. The interpretation follows in Sect. 4 with special emphasis
on the processing of the carriers of the aromatic features in the extreme irradiation
environment of PNe. This interpretation is supported by the analysis of the spectra
of compact \ion{H}{ii} regions in which the hardness of the radiation field is quite
comparable to that of the studied young PNe. 
Conclusions and implications of this work for the link between circumstellar and
interstellar PAHs are given in Sect. 5.

\section{Observations and data reduction} \label{obs}

Figure\,\ref{fig:rawspec} presents the 5-15\um\ spectra of the studied PNe in the Milky Way (left panels) and 
in the LMC and SMC (right panels). Spectra of two post-AGB objects are shown in the top (separate) panels just for 
comparison with the spectra of PNe. \object{IRAS 13416-6243} is a C-rich post-AGB object that
has been classified by \citet{pee02} as the prototype object for class C bands. Its spectrum exhibits
the 6.9\um\ feature that is seen only in the post-AGB phase  \citep{hri00}.
IRAS 13416 does not show the double-peaked spectral energy distribution that is typical in post-AGB
(cf. http://www.ncac.torun.pl/postagb; \citet{szcz07} ), which means that there is still strong
mass loss going on or that it has been stopped very recently; as a consequence, the inner radius of the envelope
is still close to the central star.
\object{MSX LMC 580} appears to have the same characteristics as IRAS 13416 with a strong 6.9\um\
feature, in particular. We therefore classify it as a genuine post-AGB object in LMC.

The ISO SWS01 spectra of the galactic post-AGB object IRAS\,13416$-$6243 and galactic PNe
(\object{Hen 3-1333}, \object{Hen 2-113}, \object{PN Vo 1}, and \object{BD+30 3639}) were reduced
using ISAP \citep{szcz97, szcz01}, while the Spitzer Space Telescope 
(SST) low-resolution InfraRed Spectrograph (IRS) spectra
(MSX LMC 580 and \object{MSX LMC 616} - acquired from programme \#3591; 
\object{SMP LMC 71}, \object{SMP LMC 79},  \object{SMP LMC 99}, and \object{SMP SMC 19} -
acquired from programme \#20443) were reduced using the
CUBISM software \citep{smi07} without the slit-loss correction function, in a 4x2 pixel aperture. Background subtraction was
achieved using an off-source spectrum in the same aperture.
The above SMP objects have a well-established evolutionary status as being planetary
nebulae \citep{sta07}, whereas MSX LMC 616 is the newly 
discovered planetary nebula in the LMC \citep{rei06}.

For the discussion of the link between aromatic dust in circumstellar and interstellar environments, we have also 
included compact \ion{H}{ii} regions in the MW and LMC. For the MW, \object{IRAS 18317-0757}, \object{IRAS 18502+0051}, 
\object{IRAS 19442+2427}, \object{IRAS 22308+5812}, and \object{IRAS 23133+6050} were selected as they have well-described 
AIB features \citep{pee02}. The ISO SWS01 spectra of the selected galactic compact \ion{H}{ii} regions were 
reduced using ISAP and are presented in the left panels in Fig.\,\ref{fig:fitHII}. For the 
LMC \ion{H}{ii} regions we have selected four sources from the SST programme \#3591
(\object{MSX LMC 1121}, \object{MSX LMC 1207}, \object{MSX LMC 1217},
and \object{MSX LMC 1798}), and reduced spectroscopic data using the CUBISM software. We have checked that these sources
are not known or newly discovered PNe in LMC \citep{rei06}, so the presence of gas emission lines in their IRS
spectra allows us to assume that they are 
genuine \ion{H}{ii} regions. Their mid-IR spectra, after processing as described in Sect.\,\ref{secfit}
are presented in the right panels in Fig.\,\ref{fig:fitHII}.



\section{Analysis of spectra using a PAH/VSG approach} \label{anal}

\subsection{The basis}
\label{bas}

Using a mathematical decomposition, \citet{rap05} and \citet{ber07} were able to extract 
in PDRs the emission spectra of PAH neutrals  (PAH$^0$) and cations (PAH$^+$),
as well as a population of carbonaceous VSGs.
The analysis of the mid-IR  (6-14\um) emission of PNe presented in this paper
is based on these results. The first step consists in composing a set of template spectra
of PAH$^0$, PAH$^+$, and VSGs.
To do this we built an average spectrum of each specie from the spectra extracted by
Rapacioli et al. and Bern\'e et al. in NGC 7023, Ced 201, and the $\rho$-Ophiucus filament.
The continuum from VSG spectra was removed since only bands are fitted (cf. Sect.\,\ref{secfit}).
Each average spectrum is then fitted using a combination of Lorentzians
and normalised so that
\begin{equation} \label{Eq}
\int_{6\mu\,m}^{14\mu\,m} I_{\nu}^{VSG}d\nu = \int_{6\mu\,m}^{14\mu\,m} I_{\nu}^{PAH^0}d\nu= 
\int_{6\mu\,m}^{14\mu\,m} I_{\nu}^{PAH^+}d\nu = 1
\end{equation}
where $I_{\nu}$ represents the specific intensity in the average spectra.
The corresponding parameters are reported in Table\,\ref{tab:temp}.

\begin{figure}
\includegraphics[width=8 cm]{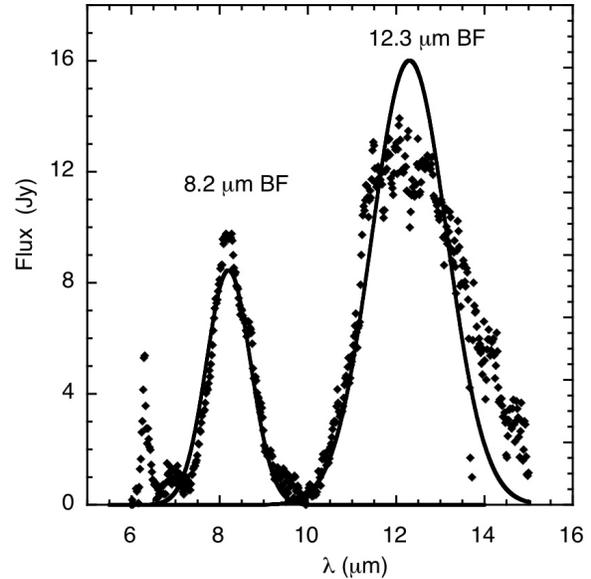}
\vspace{-0.0cm}
\caption{Spectrum after continuum subtraction of the post-AGB star IRAS\,13416-6243 (diamonds)
defined as the Type C class prototype by \citet{pee02} and template spectra for the broad features (BF)
at 8.2 and 12.3 \um\ (full line), which are used in this work.}
\vspace{-0.0cm}
\label{fig:typeC}
\end{figure}

The basis consisting of the three PDR components was found to be insufficient
for obtaining a satisfactory fit to the observations.  Broad emission features (BF) near
8 and 12\um\ are known to be prominent in post-AGB star spectra \citep{bus93, hri00, kwo01,pee02}
and a remnant of these bands could be present in PNe.
Peeters et al. have classified as Type C the objects in which the 8.2\um\ BF is observed.
We chose to use simple and necessarily approximated  Gaussian profiles with centre
at 8.2 and 12.3 \um\ (cf. Table\,\ref{tab:temp}) providing a reasonable fit to the bands observed in the post-AGB star
IRAS\,13416$-$6243, the prototype class C object from \citet[cf. Fig.\,\ref{fig:typeC}]{pee02}.

\begin{figure}[h!]
\includegraphics[width=\hsize]{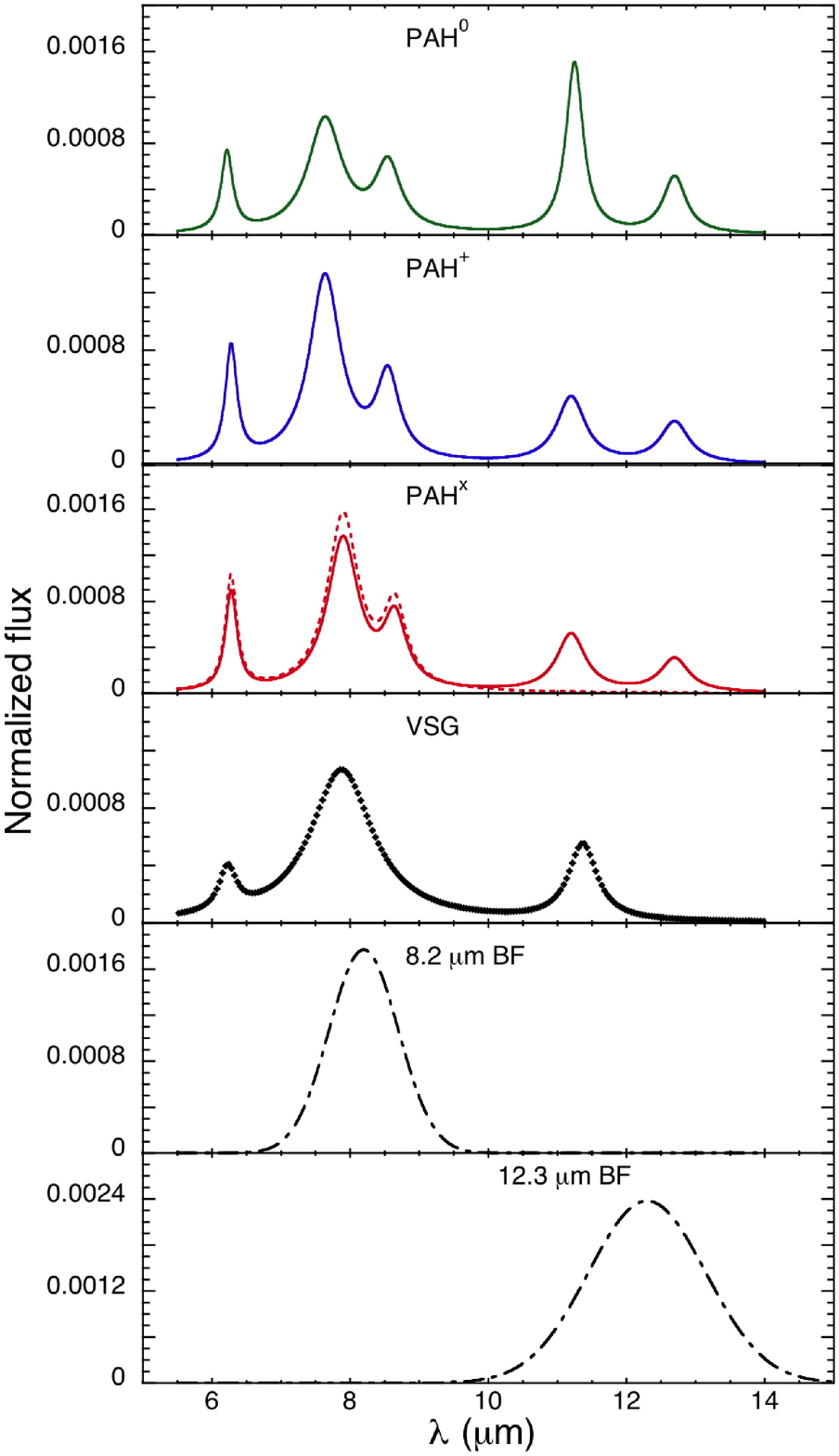}
\vspace{-0.0cm}
\caption{Summary of the template spectra used in the fit of the mid-IR
spectra. The PAH$^0$, PAH$^+$, and VSG
components are adapted from \citet{rap05, ber07}. The PAH$^x$ component is introduced
in this paper (see Sect.\,\ref{bas}) and has two possible associated spectra:
one with bands in the 10-14\um\ range (solid line), the other without (dashed line).
The 8.2 and 12.3 \um\ BFs are from Fig.\,\ref{fig:typeC}.
The flux is normalized according to Eq.\,\ref{Eq}. }
\vspace{-0.0cm}
\label{fig:specs}
\end{figure}

The ``7.7\um'' band is known to be shifted near 7.8\um\ in PNe, whereas
it occurs near 7.6\um\  in PDRs
(cf. Table\,\ref{tab:temp}). This means that a new PAH-type template spectrum has to be
introduced into our fitting procedure.
A redshifted band at  7.90\um\ was introduced with similar width and intensity
compared to the 7.64\um\  band of PAH$^+$. A redshifted counterpart
of the 8.6\um\ band was also identified (8.65\um\ instead
of 8.55\um). No counterpart was found for the 6.2 and 11.2\um\ bands.
We therefore used the characteristics of the PAH$^+$ bands
at 6.28, 11.20, and 12.70\um\ to complete the new spectrum, which is assigned to
a PAH$^x$ population. To evaluate the impact of this
assumption on the results of the fit, we built a second PAH$^x$ spectrum
with no bands in the 10-14\um\ range.
Possible candidates for the PAH$^x$ population are discussed in Sect.\,\ref{PAHproc}.

\subsection{Fitting procedure} \label{secfit}
 In extreme irradiation conditions such as in PNe, grains at thermal equilibrium
could also contribute to the mid-IR continuum. The description of this continuum would
require detailed modelling for each source. To simplify, we only consider band emission.
A continuum is therefore subtracted from the observations keeping an appropriate
level for band wing emission.  
The continuum is made of linear slopes between 6.0 and 10 \um, and
10 and 14\um. In the case of SWS spectra, the ionized  gas lines are very narrow and
can be easily subtracted.
This is more complicated for the lower resolution spectra of IRS.
However, only the [NeII] 12.8\um\ line is a concern for the fit because
of possible blend with the 12.7\um\ PAH feature. A Gaussian profile line at 12.8\um\
was added to the basis for the fit.
After continuum and line subtraction, the spectra are smoothed to the
resolution of ISOCAM data $\lambda / \Delta\lambda = 45$ \citep{bou05}.
This is justified considering that the PDR components that enter
the fit were extracted at that resolution  \citep{rap05}.
This leads to a FWHM of 0.3\um\ for the [NeII] line.
The mid-IR band emission spectra of PNe is then projected on the basis defined by the six 
template spectra from Fig.\,\ref{fig:specs} and  the ionized gas line spectrum.
The same approach was used for several compact  \ion{H}{ii} regions.
We define a \emph{Template Matrix} $T$ which contains the basis spectra.
The fit is realised by adjusting the parameters of a weight
vector $w$ so that  $\|v-wT\|^{2}$ is minimized, where  $v$  is the observed band spectrum.
This is achieved using the non-negative least squares 
(NNLS) minimization \citep{law74} which imposes $w>0$.

\subsection{Results}
\label{res}
\begin{figure*}[htp]
\includegraphics[width=8.5cm] {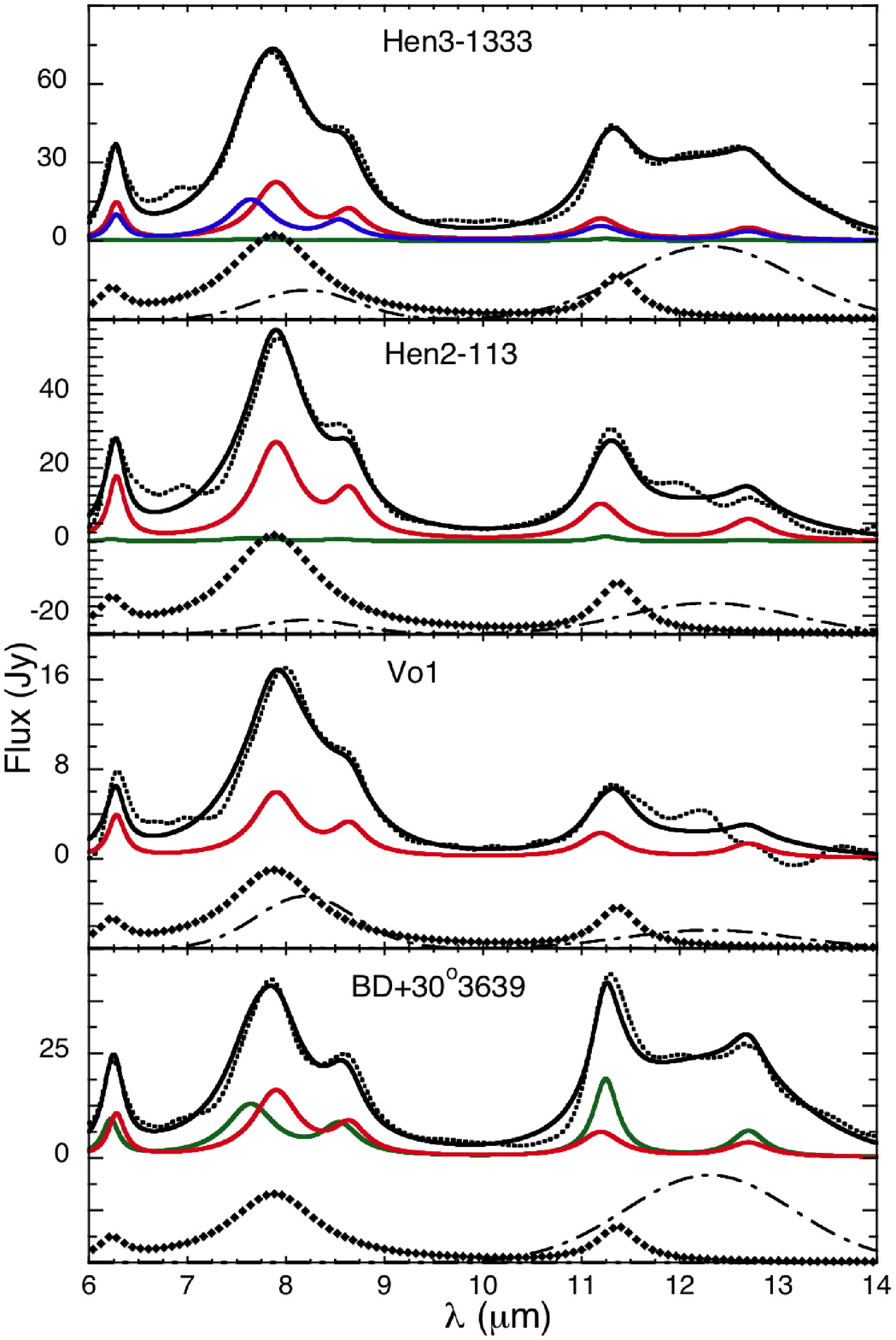}
\includegraphics[width=9cm] {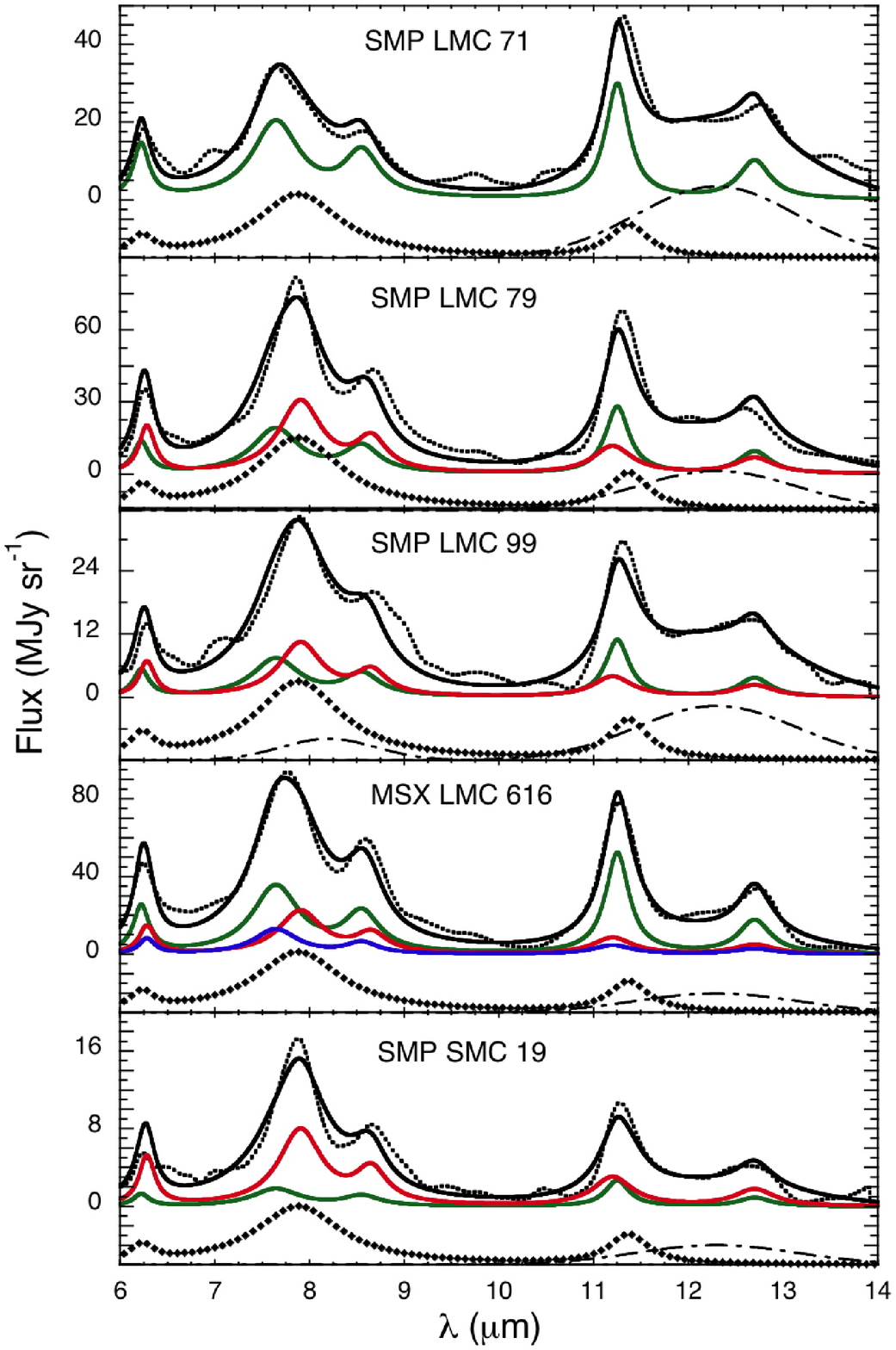}
\caption{PNe spectra after continuum subtraction and smoothing
at the resolution of $\lambda / \Delta\lambda = 45$ (dashed line) and fit using the template spectra displayed
in Fig.\,\ref{fig:specs} (solid line). The PAH components are displayed in colour lines:  PAH neutrals (green), cations (blue), and
PAH$^x$ (red). The broader components have been shifted for clarity: VSGs (diamonds), 8.2, and 12.3\um\  BFs
(dash-dot line). The ionic gas lines have been removed only in the galactic spectra recorded by SWS.}
\label{fig:fitPNe}
\end{figure*}

\begin{figure*}[htp]
\includegraphics[width=9.0 cm] {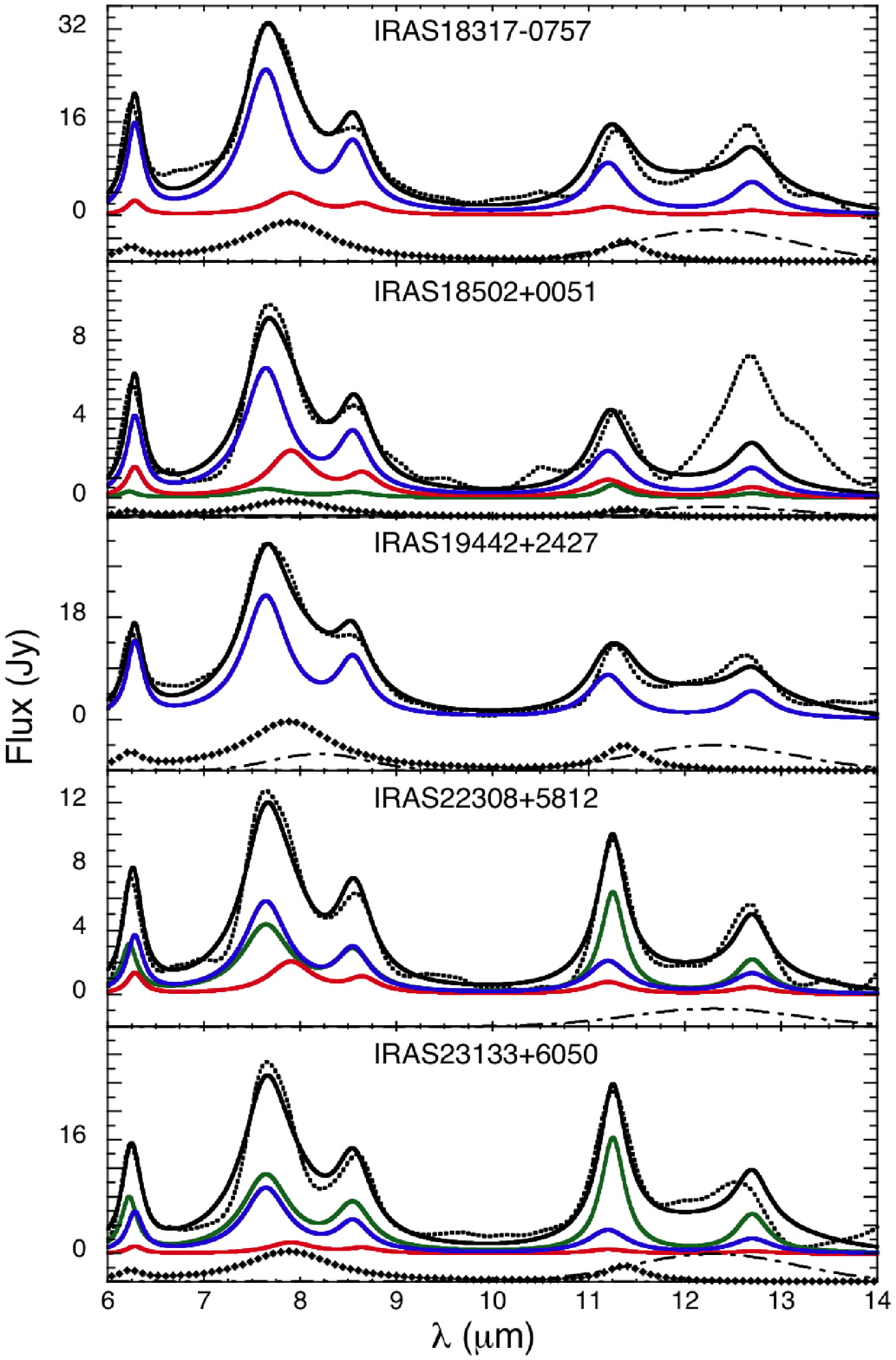}
\includegraphics[width=9.0 cm]{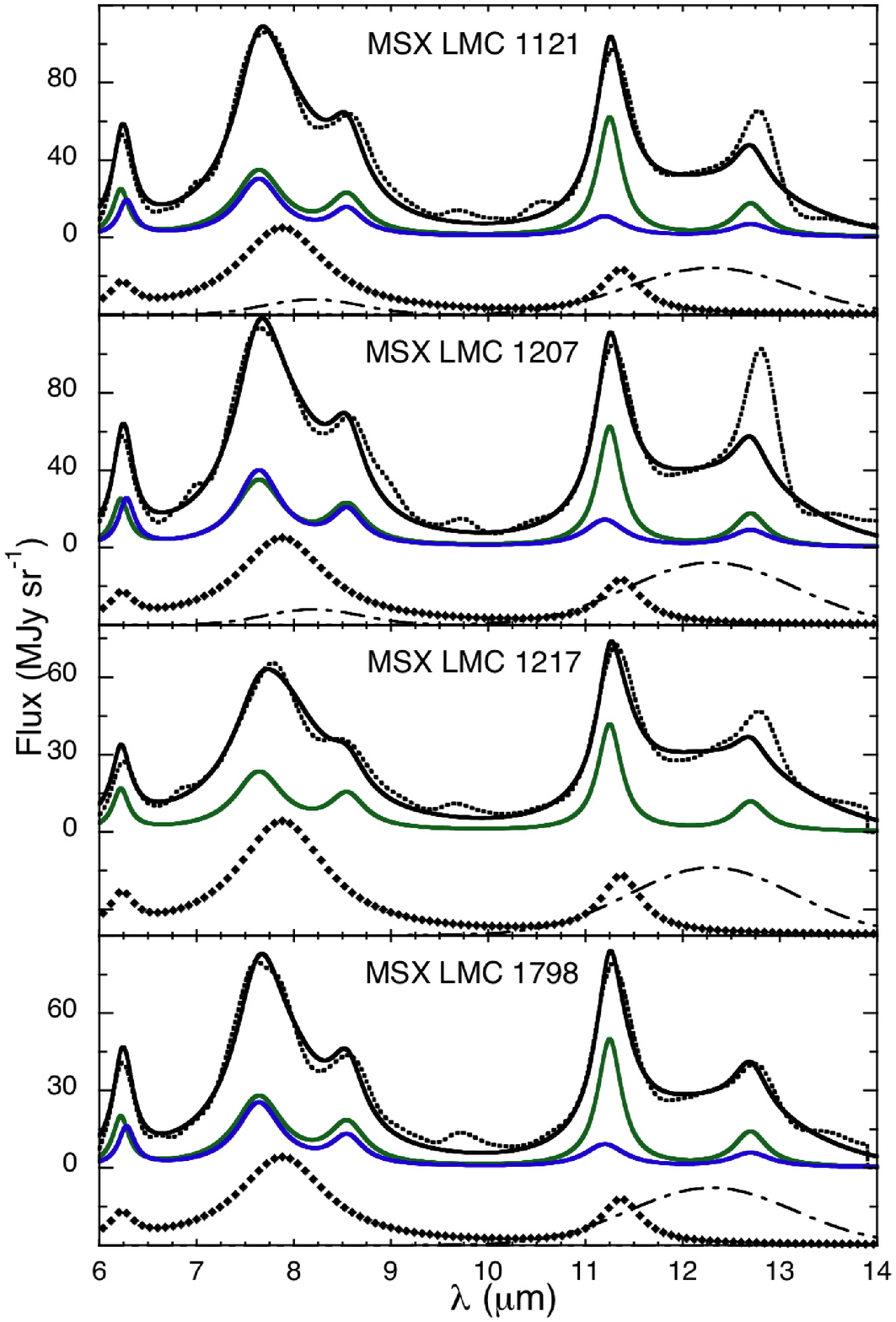}
\caption{Spectra of compact \ion{H}{ii} regions after continuum subtraction and smoothing
at the resolution of $\lambda / \Delta\lambda = 45$ (dashed line) and fit using the template spectra displayed
in Fig.\,\ref{fig:specs} (solid line). The PAH components are displayed in colour lines:  PAH neutrals (green), cations (blue), and
PAH$^x$ (red). The broader components have been shifted for clarity: VSGs (diamonds), 8.2, and 12.3\um\  BFs
(dash-dot line). The ionic gas lines have been removed only in the galactic spectra recorded by SWS. The [\ion{Ne}{ii}]
line at 12.8\um\ has been introduced as an additional component in the IRS spectra in the LMC. It is not displayed
on the plots.
}
\label{fig:fitHII}
\end{figure*}

This strategy physically
interprets the nature of the mid-IR emitters in PNe. A fit of the data with 
spline or black body continuum and Lorentzian/Drude profiles (e.g. PAHFIT, \citealt{smi07}) will
obviously give a better fit but will not give information on the properties of the emitters.
Furthermore, our fit is linear and only has 6 free parameters (7 with the [NeII] line) )
so the adjustment for one observation is instantaneous.
The results are presented in Figs.\,\ref{fig:fitPNe} and \ref{fig:fitHII} and Table\,\ref{tab1}.
The obtained fits appear quite reasonable with only one object (Hen\,3$-$1333) 
over 18 making use of the 6 dust components.
During the fitting procedure, we ran several tests. First, we found that the final results
(e.g. the relative contribution of the different components) do not depend significantly on the
exact definition of the continuum. Second, the dependence of the results on the assumption made
for the 10-14\um\ spectrum of the PAH$^x$ is given in Table\,\ref{tab1}, showing that
only slightly more PAH$^0$ emission has to be included when the PAH$^x$  do not emit
in the 10-14\um\ range.

The PAH$^x$ emission is found to be very high in galactic PNe, high in LMC/SMC PNe
(with exception of SMP LMC 71), lower but still significant  in galactic \ion{H}{ii} regions,
and undetected in LMC \ion{H}{ii} regions.
The 8.2\um\ BF is weak in all objects except in the PN Vo1. The BF features are in general
weaker in galactic  \ion{H}{ii} regions in which the spectra
are clearly affected by extinction.  To evaluate this effect, we added an extinction
factor to the fit by using the cross-section provided by \citet{wei01b} for the molecular case
(R$_V$=5.5). The column density of matter carrying the extinction is a free parameter.
We found that extinction is in the range 10-15 A$_V$ in galactic  \ion{H}{ii} regions,
a few in LMC  \ion{H}{ii} regions, and 1 at most in all PNe. Extinction effects are
therefore significant only in the former objects. By including them, we
obtained a strong increase in the PAH$^0$ component and a decrease in the VSG component,
which is consistent with extinction attenuating the 8.6\um\  and 11.3\um\ bands.
This further decrease in the VSG component  strengthens the fact that VSGs are found to be
less abundant in galactic compared to LMC \ion{H}{ii} regions.

\section{Interpretation} \label{interp}

The analysis of the mid-IR emission performed in this work shows that there are additional
mid-IR emitter components in PNe compared to the three components derived in PDRs by
\citet{rap05}. 

\subsection{Grain processing} \label{grainproc}

The first additional components are the carriers of the BFs at 8.2 and 12.3\um, which could
have associated continuum emission, but the second cannot be derived from our fitting procedure.
These are very likely to be the remnants of
the features observed in post-AGB stars \citep{hri00, pee02} where the aromatic
bands at 7.7 and 11.3\um\ are weak compared to the broad features at $\sim$8 and
$\sim$12\um. There are also additional bands, in particular the 6.9\um\ aliphatic band (cf. 
Sect.\,\ref{obs} and Fig.\,1 upper panels), suggesting that all these features are the signature
of aliphatic material that is intimately connected to the aromatic material \citep{bus93, kwo01,
got03, slo07} and strongly evolves during the transition phase from AGB to PN.
As the star becomes hotter, the material is processed by UV photochemistry.
This could lead to larger aromatic systems \citep{kwo01}.
Recent spatially resolved observations of the post-AGB object
IRAS\,22272$+$5435 \citep{got03} support a scenario of thermal processing  that leads
to an enrichment of aromatic versus aliphatic material.
The results of our fits are consistent with this destruction scenario for the carriers of the
BFs. For instance, the 8.2\um\ BF is more intense in young PNe,
the youngest object in our sample being PN Vo1 with 19\% of the mid-IR band emission in this
feature, whereas it is absent in the more evolved PN BD+30$^{\rm o}$3639.  This band is also
mainly absent  in the studied LMC PNe and  \ion{H}{ii} regions 
(cf. Figs.\,\ref{fig:fitPNe} and \ref{fig:fitHII}). In  these cases, excitation effects may play a major role
since only the hottest species are expected to emit at 8.2\um.
In particular, grains in LMC PNe may have moved farther away from the central object
before ionization becomes significant as a consequence of a larger drift between dust and gas
in lower gas density.

It has been shown in PDRs that PAHs are produced by destruction of VSGs at the surface of
molecular clouds irradiated by UV photons \citep{rap05}. VSGs
are also present in PNe with a strong emission (up to 40-50\%
of the total mid-IR band intensity both in galactic and in LMC/SMC PNe).
In galactic  \ion{H}{ii} regions, the VSG emission is significantly weaker  (cf. Sect.\,\ref{res}).
\citet{cla95} studied the emission in the 12 and 25\um\ IRAS bands in \ion{H}{ii} regions.
They concluded that the 25\um\ carriers are bound conglomerates of basic structural units and
that their destruction replenishes the 12\um\ carriers. 
This conclusion is in line with the results obtained by  Rapacioli et al. in PDRs.
That the 6-14\um\ VSG emission is higher in LMC compared to galactic
\ion{H}{ii} regions indicates that VSGs could survive longer in LMC, which reinforces the idea
that these grains are localised  farther away from the central object.

\citet{len89} underlined a dust-grain evolution scenario in PNe with an increase in the number
of dust particles and a decrease in their size with the age of the object
but the authors had difficulty concluding anything about the mechanisms
driving this evolution. Both the carriers of the BFs and VSGs can be destroyed by far-UV photons
in PNe and \ion{H}{ii} regions, and there is a trend toward decreasing abundance
of these species with the age of the PN. The chemical relation between both types
of grains, if any, is not clear though. Spatial information is required to validate
an evolutionary scenario like the one observed for the PAH/VSG transition in PDRs. 



\subsection{PAH processing} \label{PAHproc}

\begin{figure}
\includegraphics[width=\hsize]{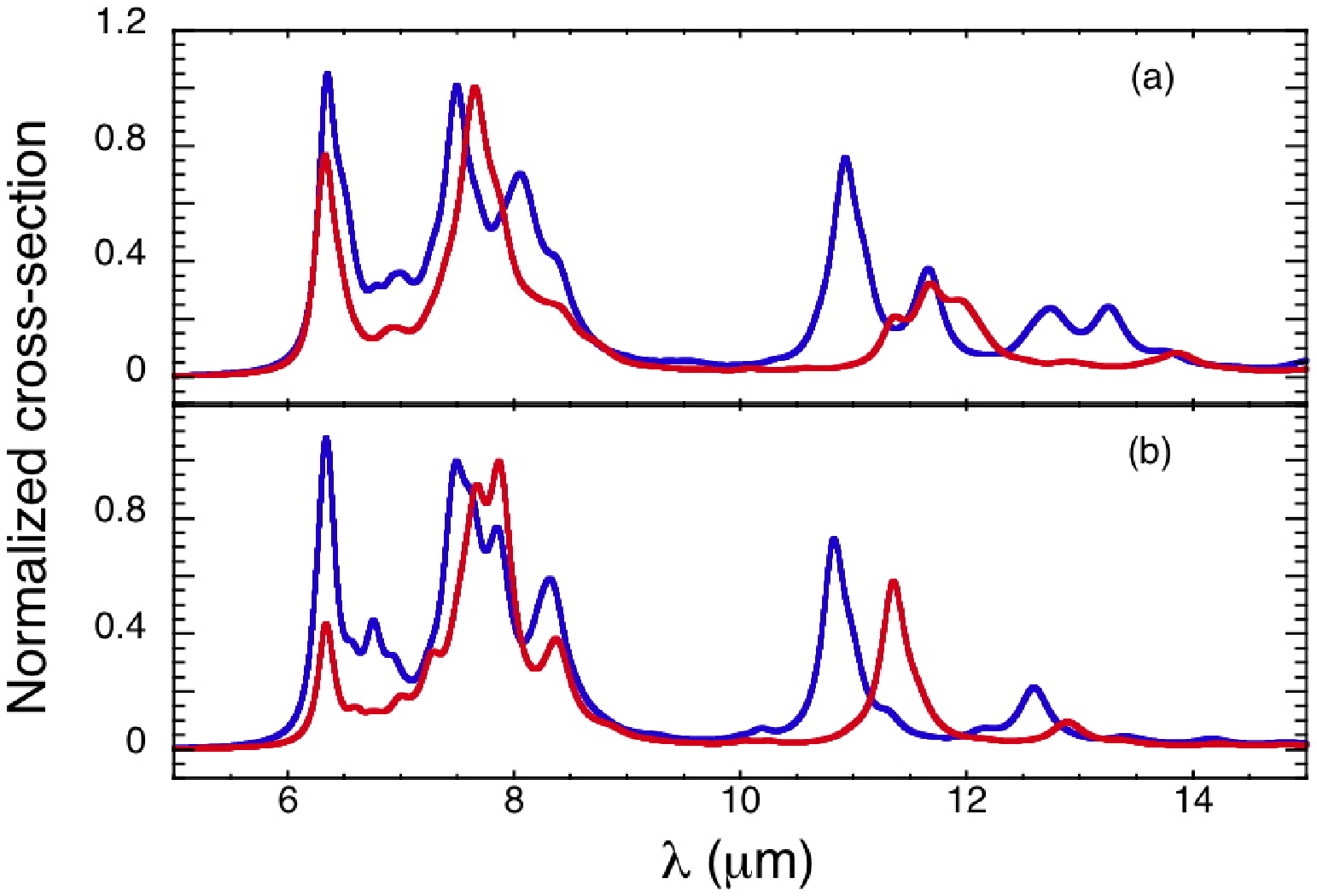}
\vspace{-0.0cm}
\caption{Theoretical IR spectra from \citet{mal07b} with cations in blue and anions in red.
(a) Average of dibenzo[bc,kl]coronene (C$_{30}$H$_{14}$),
dibenzo[bc,ef]coronene (C$_{30}$H$_{14}$), ovalene (C$_{32}$H$_{14}$),
dicoronylene (C$_{48}$H$_{20}$), and circumovalene (C$_{66}$H$_{20}$).
(b) Circumovalene (C$_{66}$H$_{20}$) only.
The spectra have been convolved with a Lorentzian line shape
of FWHM of 40 and 20\,cm$^{-1}$ for frequencies higher and smaller
than 1000\,cm$^{-1}$ (10\um), respectively. Normalization is to the maximum
of the ''7.7''\um\ feature.
} 
\label{fig:anions}
\vspace{-0.0cm}
\end{figure}

\begin{figure}
\includegraphics[width=7 cm]{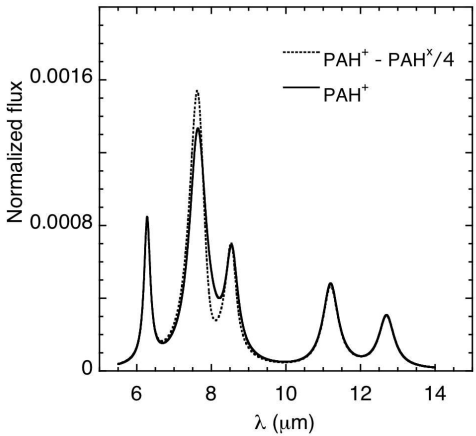}
\vspace{-0.0cm}
\caption{Template PAH$^+$ spectrum (solid line) and PAH$^+$ spectrum
after subtraction of a fraction of the PAH$^x$ spectrum (1/4). 
The flux is normalized according to Eq.\,\ref{Eq}.
} 
\label{fig_PAHx}
\vspace{-0.0cm}
\end{figure}

In our model, the 7.8\um\ band characteristic of Class B objects, essentially PNe and isolated Herbig
AeBe stars \citep{pee02}, is attributed to a PAH-type band located at 7.90\um. 
Peeters et al. propose that the class B objects reveal pristine PAHs whereas in
class A objects, which are typical of PDR-type environments, PAHs
have been significantly processed by shock waves, UV photons, and cosmic rays.
In this scenario, chemical pathways have to be found to reform the carriers of the 7.90\um\ band
in protoplanetary disks where they are also found to strongly emit \citep{ber08}.
However, as discussed in the following, our work suggests an opposite scenario in
which the class B spectra reveal more processed PAH populations.

Candidates for the 7.90\um\ band and associated PAH$^x$ spectrum were searched for in
the on-line database of the spectral properties of PAHs from  \citet{mal07b}.
We found that PAH anions might be plausible candidates since they
have spectral characteristics with the right trend (Fig.\,\ref{fig:anions}):
no shift for the 6.2\um\ band, redshift of the 7.7 and 8.6\um\ bands, and for the latter band this is better seen
in the case of the large molecule C$_{66}$H$_{20}$.
However, in the 10-14\um\ range, the strong CH out-of-plane bending mode is significantly redshifted by
about 0.5\um\ in anions compared to cations. Recent calculations on very large compact PAHs \citep{bau08}
show that this redshift tends to decrease with size. For instance, the main CH out-of-plane bending
band position was calculated to be 10.82 and 11.12\um\  for C$_{130}$H$_{28}^{+}$ and
C$_{130}$H$_{28}^{-}$, respectively.
For comparison with the observations, the theoretical band positions have to be corrected
for temperature effects, which generally lead to a redshift
of the bands with temperature  \citep{job95}. This effect has been quantified for only a few small and
medium-sized neutral PAHs; therefore, we cannot decide whether large PAH$^{-}$ or PAH$^{+}$ are better
candidates for the PAH$^{x}$ population.
On the other hand, small and medium-sized PAH$^{-}$ are excluded since they would
give a significant contribution in the 12\um\ range (cf. Fig.\,\ref{fig:anions}), which is not observed.

In their study of protoplanetary disks, \citet{ber08} conclude that PAH$^{x}$ contain larger species than
the PDR-type PAH$^{0}$ and PAH$^{+}$ populations, since they are able to survive longer under the conditions of
very high UV flux found in Herbig Ae stars.  Furthermore, the emission of PAH$^{x}$ in PDRs might be weaker
due to milder excitation conditions,  so its fingerprint might be lost in the intense PAH$^{0}$/PAH$^{+}$
spectrum (cf. Fig.\,\ref{fig_PAHx}).
PAH$^{x}$ could absorb multiple UV photons when very close to the central star, which is clearly the case
in protoplanetary disks \citep{ber08}. They could also absorb EUV photons ($h\nu > 13.6$\,eV) in \ion{H}{ii}
regions if they can survive from destruction.
In the case of PNe, it is not excluded that some dust can survive in the ionized region as previously
suggested by different authors \citep{nat81, len89, sta99, van00}.
\citet{bal96} conclude from their study of ultra-compact \ion{H}{ii} regions
that dust has to be located inside the ionized region or very close to it.
 \citet{com07} measured PAH emission in the \ion{H}{ii} region facing
the Horsehead nebula. Therefore the excitation of  PAH$^{x}$ by EUV photons is plausible.

Large anions, as well as cations, are good candidates for the PAH$^{x}$ population from the point of view
of spectroscopy. In terms of chemistry, the  PAH anion scenario is attractive since the electron density
is high in PNe and attachment by collisions could overcome detachment by UV photons
(cf. models by \citealt{bak94, wei01}).
Calculations of the PAH charge were performed for different values of the
ionization parameter \citep{szcz08} and show that PAH$^{-}$  will be present in
environments characterised by a small ionization parameter.
In LMC/SMC PNe, the lower metallicity implies a lower dust density and therefore
less attenuation of the UV field. Also, less dust implies a lower mass-loss rate during
the AGB phase and consequently lower gas and electron densities
in the PN phase. In terms of PAH charging, the combination of a higher UV flux
and a lower value of $n_{\rm e}$ will not favour a large anion
population. There is indeed a trend for the PAH$^{x}$ component to be weaker
in the LMC/SMC than in galactic PNe (cf. Fig.\,\ref{fig:fitPNe} and Table\,\ref{tab1}).
The extreme case is LMC\,71 in which there is no evidence of PAH$^{x}$.
The properties of the LMC PNe are less known than those of
galactic PNe. Still,  \citet{lei06} assume for LMC\,71 a value of $n_{\rm e}$ of 5000\,cm$^{-3}$,
relatively higher than the values   3100, 3600, and 2300\,cm$^{-3}$ in
LMC\,79, LMC\,99, and SMC\,19, respectively.
According to these authors, LMC\,71 also has the highest excitation class : T$_{eff}\sim$80000\,K
and Log\,L/L$_{}{\sun} < 4.27$ compared to T$_{eff}\sim$33000\,K and Log\,L/L$_{}{\sun} < 3.79$
in LMC\,79 \citep{vil03}. This could explain why PAH$^{-}$ cannot survive in LMC\,71.
For the galactic \ion{H}{ii} regions, the value of $n_{\rm e}$
is about 1500, 1200, 770, and 480\,cm$^{-3}$ in IRAS\,18317$-$0757,
IRAS\,18502$+$0051, IRAS\,22308$+$5812, and IRAS\,23133$+$6050, respectively
\citep{rud06, mar02}. It is about two orders of magnitude lower than in PNe in
which values in the range from about 50\,000 up to 100\,000\,cm$^{-3}$
have been determined for Hen\,3$-$1333, Hen\,2$-$113, and Vo1 \citep{dem97, men90}, and a value
of 21\,000\,cm$^{-3}$ in BD$+$30$^o$3639, the most evolved PN in our sample \citep{med06}.
A detailed modelling of the PAH charge in each of these regions is required, but this
implies a good knowledge of the characteristics of each source.

\section{Conclusion} \label{conc}

The analysis of the mid-IR spectra in different objects
has led to useful object classifications and some evolutionary scenario for PAH species
 \citep{hon01, pee02, van04, slo07}. 
However, these studies are limited by the difficulty interpreting the observed spectra as
the sum of individual contributions tracing different chemical populations. 
The present analysis provides a more comprehensive
view over the 6-14\um\ range based on a previous analysis of the mid-IR spectra of PDRs
\citep{ber07}.  We propose to rationalize the interpretation of the ``7.7"\um\ feature
in evolved stars with the following evolutionary scenario.
In cool and less evolved stars (post-AGB stage), the emission is dominated by aliphatic rich
material as suggested by different authors. In hotter star environments, this material
has been processed and aromatic material dominates the emission.
The destruction of VSGs to produce free PAHs is operating as observed in PDRs
\citep{rap05, ber07}, but there is an additional step in PAH processing that
results in the selective emission of very large PAHs. The latter PAH$^x$ population
is only observed in extreme irradiation environments, most likely due to the combined
effect of the destruction of the smallest PAHs with an increase in the temperature of
PAH$^x$ due to the high UV flux and the presence of EUV photons.
Spatial information would be required, though, before
concluding that the emitting PAH$^x$ are located in \ion{H}{ii} regions, and
this is indeed one of  the goals of the SPECHII programme currently
running on the Spitzer Space Telescope.
Interestingly, the PAH$^x$ species are also present in most LMC/SMC PNe,
whereas they seem to be absent in LMC  \ion{H}{ii} regions. This provides an observational
diagnostic to differentiate both types of regions, which
might be more reliable than the one usually used, e.g. the \citet{buc06}
2MASS-MSX colour classification system (cf. for instance \citealt{kas08}).
Finally, a scenario in which these PAH$^{x}$ are negatively charged
is plausible but lack quantitative studies. If one could come to some conclusion on this point,
then  the PAH$^{x}$ emission could be used to constrain the physical parameters
(UV field, electronic density).
Our goal is not to decide on this point but to lay the groundwork for more detailed studies,
in particular involving a larger sample of objects and a better description of their
physical parameters.


This work provides further support to the idea that evolved carbonaceous stars are the PAH nursery.
The same scenario seems to apply in the LMC/SMC, indicating that the change
in metallicity has not strongly affected the condensation sequence of aromatic dust,
contrary to what has been suggested by some authors \citep{spe06}.



\begin{acknowledgements}
  
We thank the referee and the editor for useful comments that improved the
manuscript. 
R.Sz. ackowledges support from grant N203 019 31/2874 of the Science
and High Education Ministry of Poland. Support from the French national
programme, ÒPhysique et Chimie du Milieu InterstellaireÓ, is also 
acknowledged.

\end{acknowledgements}

\bibliographystyle{aa}
\bibliography{biblioCJ}

\Online
\appendix
\section{Tables}
\begin{table*}
\caption{Positions, full widths at half maximum (FWHM), and peak intensities (relative to the 7.7 $\mu$m band)
of the bands of the template spectra (cf. Fig.\,\ref{fig:specs}). Lorentzian shapes have been used except when specified.
For PAH$^x$, two cases are considered for the 10-14\um\ range, one with bands similar to PAH$^+$,
the other in italics without bands.}
\label{tab:temp}
\begin{center}
\begin{tabular}{lccccccccccc}
\hline \hline
 
  & \multicolumn{3}{c}{6.2} &  & \multicolumn{3}{c}{7.7} &  &\multicolumn{3}{c}{8.6} \\
\cline{2-4} \cline{6-8} \cline{10-12}  \vspace{-0.25cm} \\
 & Pos. &  FWHM & Int. &  & Pos. &  FWHM & Int. &  & Pos. &  FWHM & Int. \\
\hline
PAH$^{0}$  & 6.22 & 0.17 & 0.75 &  & 7.64 & 0.60 & 1.0 &  & 8.55 & 0.45 & 0.57 \\
PAH$^+$    & 6.28 & 0.20 & 0.65 &  & 7.64 & 0.55 & 1.0 &  & 8.55 & 0.40 & 0.45 \\
PAH$^x$    & 6.28 & 0.20 & 0.65 &  & 7.90 & 0.55 & 1.0 &  & 8.65 & 0.40 & 0.45  \\
                    & {\it 6.28} & {\it 0.20} & {\it 0.65} &  & {\it 7.90} & {\it 0.55} & {\it 1.0} &  & {\it 8.65} & {\it 0.40} & {\it 0.45}  \\
VSG       & 6.23 & 0.30 & 0.64 &  & 7.88 & 1.10 & 1.0 &  &  -   &  -   &  -   \\
$^\dagger$8.2\um\  BF      &      &      &      &  & 8.20 & 1.18 &     &  &      &      &       \\
\hline
  & \multicolumn{3}{c}{11.3} &  & \multicolumn{3}{c}{12.7} &  &\multicolumn{3}{c}{}  \\
\cline{2-4} \cline{6-8}    \vspace{-0.25cm}\\
 & Pos. &  FWHM & Int. &  & Pos. &  FWHM & Int. &  &  &   &  \\
\hline
PAH$^{0}$  & 11.25 & 0.3 & 1.60&   & 12.70 & 0.4 & 0.54 \\
PAH$^+$    & 11.20 & 0.5 & 0.38&   & 12.70 & 0.5 & 0.22 \\
PAH$^x$    & 11.20 & 0.5 & 0.38 &   & 12.70 &  0.5     &  0.22   \\
   & - & - & {\it 0} &   - &  -   &   - &  {\it 0}    &  \\
VSG      & 11.37 & 0.5 & 0.32&   &       &      &  \\
$^\dagger$12.3\um\  BF       &      &      &     &   & 12.30 & 2.00 &  \\
\hline
\multicolumn{12}{l}{$^\dagger$ Parameters for a Gaussian}
\end{tabular}
\end{center}
\end{table*}

\begin{table*}[ht!]
\caption{Integrated flux in the observed PNe for each component of the fit:  PAH$^0$, PAH$^+$,  PAH$^x$,  VSG,
8.2, and 12.3\um\  BFs (cf. spectra in Figs.\,\ref{fig:fitPNe} and
\ref{fig:fitHII}). Numbers in parentheses show the relative contribution in \%\ of the different ionization states of PAHs.
For each object, the two templates of the PAH$^x$ spectrum have been used (cf. Table\,\ref{tab:temp}).
[ArII] means a line at 6.99\um, [ArIII] at 8.99\um, [SIV] at 10.51\um, [NeII] at 12.81\um, and
* means that the presence of the line emission was determined from the fitting procedure.}
\label{tab1}
\begin{center}
\begin{tabular}{llllllll}
\hline \hline
\noalign{\smallskip}
 
  Object   & \multicolumn{6}{c}{Normalized 6-14\um\ integrated flux}&     Ionized \\
  & \multicolumn{6}{c} {(percentage of PAH flux)} &    gas lines  \\
\cline{2-7} \noalign{\smallskip}
    & PAH$^0$ & PAH$^+$ & PAH$^x$ &  VSG & 8.2\um\ BF & 12.3\um\ BF & \\
 \hline
{\bf Galactic PNe} \\
Hen 3-1333    &  0.01  (0) &   0.15  (42)   &   0.21 (58) &   0.38 &     0.09 &   0.16 &  [NeII]  \\
  &  {\it 0.11 (27)} &   {\it 0.08  (20)}   &   {\it 0.21 (53)} &   {\it 0.36} &    {\it 0.07}  & {\it 0.17}   &  \\
Hen 2-113       &  0.02 (5) &   0   &   0.39 (95) &   0.48  &    0.04 &    0.07  &  weak [ArII], weak [ArIII], [NeII] \\
& {\it  0.13 (33)} &   {\it 0}   &   {\it 0.26 (67)} &   {\it 0.47}  &    {\it 0.05} &    {\it 0.09}  & \\
Vo1 &  0 &   0   &  0.31 (100)  & 0.45  &  0.19  &   0.05  & [NeII] \\
 &  {\it 0.01  (4)} &   {\it 0}   &   {\it 0.24 (96)} &   {\it 0.51} &    {\it 0.17} &    {\it 0.07} & \\
BD+30$^{\rm o}$3639  &  0.26 (52) &   0    &   0.24 (48) &   0.31 &  0   &    0.19  &  [ArII], weak [ArIII], [NeII] \\
 &  {\it 0.33 (65)} &   {\it 0}    &   {\it 0.18 (35)} &   {\it 0.29} &  {\it  0}   &   {\it  0.20} &   \\
\hline
{\bf LMC/SMC PNe} \\
SMP LMC 71 & 0.46 (100)  & 0  &    0 &   0.35  &   0  &   0.19   & -\\
                     & {\it 0.46 (100)}  & {\it 0}  &   {\it 0} &   {\it 0.35}  & {\it 0}   &   {\it 0.19}   &  \\
SMP LMC 79 &  0.25  (45) &   0    &  0.30 (55) &   0.36 & 0 &    0.09    & -\\
                      &  {\it 0.33  (62)} &  {\it 0}  &  {\it 0.20 (38)}  &   {\it 0.37}  &   {\it 0} &  {\it 0.10}   & \\
SMP LMC 99 &  0.20 (48)   &  0   &  0.22 (52)  &  0.38  &  0.07 &  0.13   & -\\
                         &   {\it 0.27 (69)} &    {\it 0}  & {\it  0.12 (31) } &  {\it 0.40}  &   {\it 0.07}  & {\it  0.14}   & \\
MSX LMC 616  & 0.37 (57) &   0.10 (15)  &   0.18 (28)  &  0.31  &   0  &  0.04 &  weak [NeII]* \\
                             & {\it 0.44  (69)} &   {\it 0.06 (9)} &  {\it 0.14 (22)}  &  {\it 0.31}  & {\it 0}  &   {\it 0.05} &   \\
SMP SMC 19 & 0.13 (24) &   0   &  0.42 (76)  &  0.39  & 0 &    0.06   & weak [NeII]*\\
                        &  {\it 0.25 (46)} &  {\it 0}     &  {\it 0.29  (54)} &  {\it  0.38}  & {\it 0}  &   {\it 0.08}  & \\
\hline
{\bf Galactic \ion{H}{ii} regions} \\
IRAS18317-0757 &  0  &  0.62 (87)   &  0.09 (13) &  0.21  &   0  &   0.08  & [ArII], [ArIII], weak [SIV], [NeII]\\
 &   {\it 0}  &    {\it 0.62 (85)} &  {\it 0.11 (15)}  &  {\it 0.18}   &  {\it 0}   &     {\it 0.09}   &   \\
IRAS18502+0051 &  0.05 (6) &  0.61 (70)  &  0.21 (24)  &   0.10  &   0  &   0.03  & [ArII], [ArIII], weak [SIV], [NeII]\\
                                 &     {\it 0.14  (16)}  &  {\it 0.55 (62)} &  {\it 0.19 (22)}  &    {\it 0.07}  & {\it 0}    &  {\it 0.05} &   \\
IRAS19442+2427 & 0  &  0.58 (100) &  0  &  0.29  &  0.06   &  0.07   & [ArII], [NeII]\\
                                  & {\it 0} &  {\it 0.56  (89) }  &  {\it 0.07 (11)} & {\it 0.26}   &  {\it 0.04}  &  {\it 0.07}  & \\
IRAS22308+5812 & 0.38 (42) &  0.39 (43)   & 0.13 (15) &   0.06  & 0    &   0.04 & [ArII], [ArIII], [NeII]\\
 & {\it 0.44 (48)} &  {\it 0.35 (39)}  &  {\it 0.12 (13)} &  {\it 0.04} & {\it 0}    &   {\it 0.05}  & \\
IRAS23133+6050 & 0.44 (57) &  0.28 (36)   &  0.05  (7) &  0.16  &  0  &    0.07  & [ArII], [ArIII], weak [SIV], [NeII]\\
 & {\it 0.45 (59)} &  {\it 0.28 (37)} &  {\it 0.03 (4)} &  {\it 0.17 } &  {\it 0}  &    {\it 0.07}  & \\
\hline
{\bf LMC \ion{H}{ii} regions}\\
MSX LMC 1121 &  0.31 (61) &   0.20  (39)   &   0  &   0.36 &    0.04 &    0.09 & [NeII]* \\
MSX LMC 1207 & 0.28 (54) &   0.24 (46)   &   0  &   0.33 &    0.04 &    0.11 & [NeII] *\\
MSX LMC 1217 &  0.31  (100) &   0    &   0  &   0.54 &  0   &    0.15 & [NeII]* \\
MSX LMC 1798 &  0.32 (60) &   0.21 (40)   &   0  &  0.36 &   0  &   0.11  & -  \\
\hline

\end{tabular}
\end{center}
\end{table*}

\end{document}